%% file: main.tex
\documentclass[runningheads]{llncs}
\usepackage{graphicx}
\usepackage{booktabs} 
 \usepackage{array,multirow}

\makeatletter

\title{Exploiting Graph Structured Cross-Domain Representation for Multi-Domain Recommendation}

\author{Alejandro Ariza-Casabona\inst{1}
\orcidID{0000-0002-3388-2316} \and
Bartlomiej Twardowski\inst{2,3}\orcidID{0000-0003-2117-8679}\and
Tri Kurniawan Wijaya\inst{3}\orcidID{0000-0002-8916-7442}
}

\authorrunning{A. Ariza-Casabona et al.}

\titlerunning{Graph Structured Cross-Domain Representation}

\institute{Universitat de Barcelona, Barcelona, Spain\\ \email{alejandro.ariza14@ub.edu}\and
Computer Vision Center, UAB, Barcelona, Spain\\ \email{btwardowski@cvc.uab.es}\and Huawei Ireland Research Center, Dublin, Ireland\\ \email{\{bartlomiej.twardowski, tri.kurniawan.wijaya\}@huawei.com}}

\usepackage{bibentry}
\usepackage{amssymb}
\usepackage{amsmath}

\newcommand{\minisection}[1]{\vspace{0.1in} \noindent {\bf #1}}
\newcommand{\minisec}[1]{\vspace{0.1in} \noindent {\bf \emph{#1}}}
\newcommand{\minipara}[1]{\vspace{0.1in} \noindent {\it #1}}
\usepackage{xcolor}

\begin{document}

\maketitle

\begin{abstract}

Multi-domain recommender systems benefit from cross-domain representation learning and positive knowledge transfer. Both can be achieved by introducing a specific modeling of input data (i.e. disjoint history) or trying dedicated training regimes. At the same time, treating domains as separate input sources becomes a limitation as it does not capture the interplay that naturally exists between domains.
In this work, we efficiently learn multi-domain representation of sequential users' interactions using graph neural networks.
We use temporal intra- and inter-domain interactions as contextual information for our method called MAGRec (short for \textit{\textbf{M}ulti-dom\textbf{A}in \textbf{G}raph-based \textbf{Rec}ommender}). To better capture all relations in a multi-domain setting, we learn two graph-based sequential representations simultaneously: domain-guided for recent user interest, and general for long-term interest. 
This approach helps to mitigate the negative knowledge transfer problem from multiple domains and improve overall representation. 
We perform experiments on publicly available datasets in different scenarios where MAGRec consistently outperforms state-of-the-art methods.
Furthermore, we provide an ablation study and discuss further extensions of our method.


\end{abstract}

\input{intro}
\input{related_work}
\input{methodology}

\input{discussion}
\input{conclusions}

\minisection{Acknowledgements}
This work was partially supported by the FairTransNLP-Language Project (MCIN/AEI/10.13039/501100011033/FEDER,UE).

\bibliography{ecir23}
\bibliographystyle{splncs04}

\end{document}

%% file: intro.tex
\section{Introduction}

Recommender systems are introduced to solve 
the task of quickly retrieving the most suitable items from large catalogs to the corresponding users. The complexity of the task comes not only from the huge amount of information that already subdivides the recommendation problem into three stages (matching, ranking and re-ranking) \cite{dl4rec_survey}, but also from the multi-objective minimization interest in multi-stakeholder platforms (accuracy, novelty, fairness, explainability, business income, etc.) \cite{multistakeholder_survey} and the different types of recommendation scenarios (sequential, session-based, social, cross-domain, multi-domain, etc.).

Whilst most recommender systems focus on a single domain recommendation (SDR) problem, it is increasingly common to find large-scale commercial platforms containing products from multiple domains in which the user/item set reaches a certain overlapping degree. These multiple domains may correspond to different advertisement types or product categories that present certain domain commonalities but still require specialized promotion strategies or models for effective profitability. Unfortunately, having a single “naive” model to serve all domains may achieve subpar results. 
As a consequence, several models were proposed to tackle cross-domain recommendation (CDR) with the aim of transferring knowledge from a source domain to one or multiple target domains \cite{conet,emcdr,minet}.

The main problem that appears with CDR is the fact that each domain requires separate training or fine-tuning of a different model per each set of source-target domains, but large-scale commercial platforms may consider a single model to improve performance for all domains simultaneously. This challenge has been recently introduced as Multi-Domain Recommendation (MDR) \cite{star} in which the distinct domain data distributions caused by different user behavioral patterns are modelled altogether. 
An advantage of MDR systems is that they make use of all available data for training unlike previous alternatives, resulting in positive outcomes for low-resource domains and optimal performance if shared information among domains is properly exploited.

As previously explored, one approach to improve learning on multiple domains is to use multi-task learning (MTL) models by considering each domain as a different task \cite{sharedbottom,mmoe,mulann}. However, MTL models are built to model different label spaces rather than partially distinct input data distributions that characterize the MDR problem. That is the reason why these models incorporate separate task-specific output layers and share the bottom input representational layers, leading to a poor ability to represent domain commonalities in the label space. 
To overcome MTL limitations, existing studies \cite{mamdr,star} have focused on adding domain-shared knowledge to the classification block of the recommender system, and defining new training strategies to avoid the domain conflict and domain overfitting problems.
However, little to no effort has been made on properly modeling the available multi-domain structural information from the past user behavior history. Existing approaches ignore this information and rely on a simple domain indicator.
Therefore, the input modeling capabilities are limited and excessive time-consuming training strategies are necessary to achieve a successful convergence.

In order to fully exploit domain interest fluctuations and specific data distributions, we combine the power of Sequence-Aware Recommender Systems and Graph Neural Networks (GNN) into a new model named MAGRec for MDR. MAGRec receives multi-domain graph representations of historical user interactions performed in multiple domains, and processes them using a two-branch network architecture. On the one hand, one branch focuses on the user's most recent interest via domain-aware message passing through the sequential graph. On the other hand, the second branch tries to create a contextualized global user representation via graph structure learning and local pooling operations. Finally, this rich representation of the input is combined and fed to the classification network that models the MDR task. Given that our focus lies on the input representation modeling and its ignored presence in the MDR literature, unlike previous works that adapt the classification network and training strategies, our model only uses a single fully connected network (FCN) as the classification block. This approach also benefits those cases where the number of domains is very large and both training times and model parameters reach unacceptable limits. Nonetheless, it is important to note that our domain-aware graph modeling network could be combined with previous state-of-the-art adopted strategies, which is left for future work.

The main contributions of this work can be summarized as follows:

\begin{itemize}
    \item We propose a new MDR model: MAGRec. The use of graph neural networks to capture domain relationships from past user interactions in multiple domains, together with the integration of global and recent user interest with domain-shared Graph Structure Learning, provides a faster alternative to MDR that puts the focus on the input modeling stage of the model.
    \item We explore different input representations from a “naive” to a more complex multi-domain sequential representation and test them on multiple sequential and graph-based recommender systems.
    \item Extensive experiments on different MDR scenarios and models, including: single-domain sequential CTR prediction models, MTL recommendation models, and MDR strategies show the consistent viability of our model, and it clearly opens new directions in MDR research such as the combination of input representation capabilities with task modeling strategies.
\end{itemize}

%% file: related_work.tex
\section{Related Work}

Sequence-aware and graph recommender systems are commonly applied for the SDR problem. In our work, we propose an approach based on recent advancement from latest research in those fields to tackle the MDR problem, which we describe below. 

\minisection{Sequence-Aware CTR Prediction} 
The path that single-domain CTR prediction models have followed over the past few years go from shallow models \cite{mcmahan2013ad} to complex deep learning architectures. The later are trying to extract more complex patterns from users' behavior by adding feature interaction modeling strategies such as cross-connections \cite{dcn}, improving numerical features representation \cite{autodis} or exploiting sequential patterns in the past user's behavior \cite{dien,din}, just to name a few. However, it was a sequential modeling that improved recommender systems in many domains with the introduction of recurrent architectures \cite{gru4rec}, convolutional networks \cite{caser} and attention modules \cite{sasrec,narm}. Therefore, they have been adopted for CTR prediction task as well. However, these techniques remain unexplored in MDR where domain sequential interest fluctuations must be correctly accounted for.

An important remark from recent sequence-aware models is the necessity to properly integrate the long- and short-term user interest representations \cite{surge,sdm,sim}. In this work, we integrate domain contextualization for both short- and long-term user modeling. The former is achieved via a weighted edge message passing mechanism, and the latter via domain-aware graph structure learning and local pooling strategies.

\minisection{Graph Neural Networks for Recommendation} 
The use of GNNs has notably improved sequential recommendation \cite{surge,fgnn,srgnn}. The importance of structure modeling has even been addressed with hypergraph representations to account for higher-order node relations. \cite{dhcn} proved the beneficial effects of combining different input graph representations such as hypergraphs and reduced line graphs for session-based user representation. Furthermore, \cite{surge} support previous findings regarding the necessity of extracting global user interest with graph clustering (for noise removal) and properly fuse it with the most recent user interests for more accurate recommendations.

In our work, we use GNNs to model intra- and inter-domain connectivity patterns from a graph-based multi-domain user history representation by exploiting item node features and domain-related edge features.

\minisection{Multi-Domain Recommendation} Unlike CDR, where source and target domains are clearly defined and most benchmarks assess a knowledge transfer between domains, MDR aims at improving performance on all domains simultaneously with, preferably, a single model to reduce computational complexity.

On the one hand, having a single traditional model without domain knowledge and sharing all parameters to serve all domains may put a huge burden on the generalizability-specialty trade-off of the downstream task. On the other hand, having a separate model per domain is unacceptable in platforms with many domains, due to the number of trainable parameters and lack of training data for certain domains. A middle ground for CDR and MDR can be found in \cite{herograph,dasl,ppgn}. \cite{dasl} proposed a sequential modeling of the input with dual embedding and dual attention mechanisms. However, generalizing this work for more than two domains is not trivial. 
\cite{herograph,ppgn} created a cross-domain graph representation but ignored the sequential connections that are important for inter-domain interest flow. Moreover, generalization to an arbitrary number of domains is barely explored.

Alternatively, other approaches to MDR included several MTL architectures in order to model each domain as a separate task \cite{mmoeex,sharedbottom,mmoe,ple}. All of them have task-shared input representation network i.e., expert networks, and multiple task-specific networks for a final prediction denoted as tower networks by MMoE. Nonetheless, as previously mentioned, the MDR problem differs from MTL in the fact that the task/label space is the same and face different input distributions from multiple domains. Consequently, \cite{star} focused their efforts on building a task-modeling network with star topology that could model all domains as a shared task called STAR.
It proved the importance of domain contextualization with the use of a partitioned normalization per domain and an auxiliary network.
Despite improving MDR performance, STAR still suffers from the domain conflict problem 
and it is prone to overfit on sparse domains as stated in \cite{mamdr}. To overcome those limitations, \cite{mamdr} introduced MAMDR as an attempt to adapt the MDR training strategy to include a domain negotiation and a domain regularization approach. Nevertheless, as these MDR approaches were built on top of MTL strategies and despite STAR supporting the evidence that a simple domain indicator is important contextual information in these scenarios, they pay little attention on how to model domain interconnections and structural information from the input. 

In this work, we aim to prove the relevance of structured multi-domain representation in MDR problems. Note that \cite{mamdr,star} are complementary to this work and could potentially be used in combination with our model to overcome task modelling limitations on many-domain scenarios, thus, boosting the performance.

%% file: methodology.tex
\section{Methodology}

In this section, we provide a brief introduction to the MDR problem and our proposed two-branch architecture. Furthermore, we detail how domain information is exploited by each branch to leverage recent and global user interest representations.

\minisection{Problem Formulation} In traditional recommendation scenarios, it is common to have a set of users and a set of items, denoted by $\mathcal{U}$ and $\mathcal{I}$ respectively. In addition, we have a set of user interactions denoted by $\mathcal{B}$. In the single domain setting, each interaction $b \in \mathcal{B}$ is a tuple $(u, i, t, y)$, where $u\in \mathcal{U}$, $i\in \mathcal{I}$, $t$ is the timestamp of the interaction and $y$ is the user action (e.g.~was there a click or not in CTR prediction).  In the MDR setting, each tuple must further contain the domain $d \in \mathcal{D}$ in which that interaction took place, i.e. $b = (d, u, i, t, y)$.

The goal in a conventional single-domain setting is to predict the action $y$ a user $u$ will take for a candidate next item $i_{k}$ in a domain $d$, given their previous behaviors as a sequence in that particular domain $\mathcal{B}^{d}_{u} = [b^{d}_{u,1}, b^{d}_{u,2}, ..., b^{d}_{u,N}]$. However, when there are $|\mathcal{D}|$ different domains that need to be taken into account, some domains may encounter partial or total overlap in their user and/or item sets.
Consequently, user history $\mathcal{B}_{u}$ can be interpreted in multiple ways depending on how domain data is aggregated and which interactions are considered. In order to account for more complex representations to be fed into the model, $\mathcal{B}_u$ is transformed into the corresponding graph representation $\mathcal{G}_u = \{\mathcal{V}_u, \mathcal{E}_u\}$, where $\mathcal{V}_u$ is the set of vertices/items, and $\mathcal{E}_u$ is the set of edges containing intra- and, potentially, inter-domain sequential connections. This transformation is data dependent, see \emph{Experiments} section.

\minisection{Proposed Architecture} Our proposed MAGRec model is primarily built to maximize input representation capabilities for the multi-domain recommendation problem. Its architecture is illustrated in Figure \ref{fig:architecture}. As previously mentioned, the input includes the user $u$, the selected user history representation $\mathcal{G}_u$, the candidate item $i_k$ and the domain to which the candidate item belongs $d_k$. Note that the graph contains both node and edge features. Specifically, node features are represented by item embeddings and edge attributes correspond to the source and target domains in the sense of a directional connection. Next, a brief explanation of each block of our architecture is presented:

\begin{figure}[tb!]
    \centering
    \includegraphics[width=.7\linewidth]{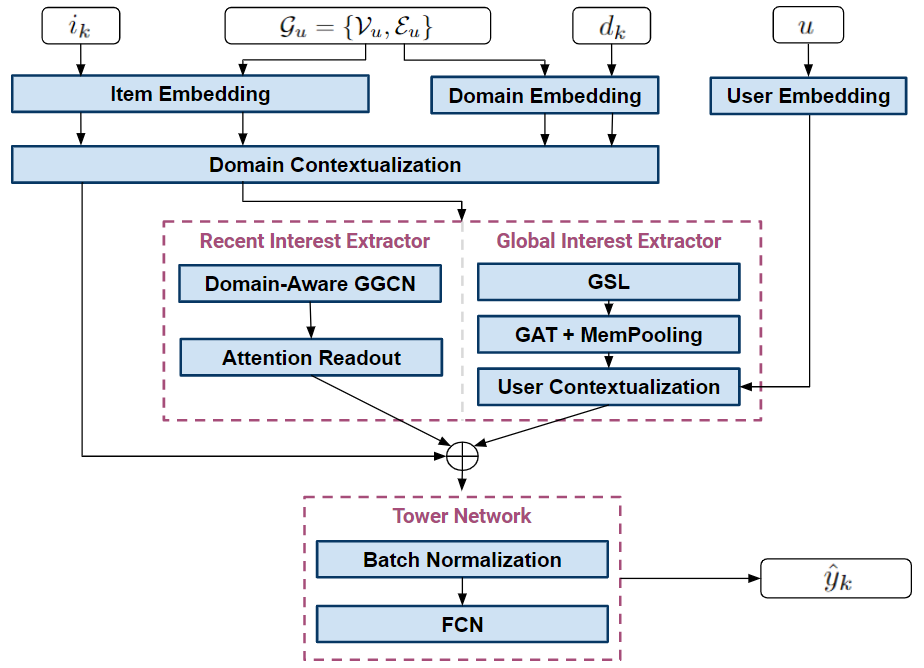}
    \caption{Illustration of MAGRec. MAGRec extracts both recent and global user interests from a contextualized multi-domain user history graph representation. The recent interest extractor learns cross-domain sequential patterns from intra- and inter-domain interactions focused on the last user-item interaction. The global interest extractor learns item correlations and their affinity to long-term user core preferences via graph structure learning and local graph pooling techniques.
    }
    \label{fig:architecture}
\end{figure}

\minisec{Embedding} An initial embedding layer is applied to the items, users and domains in order to obtain their latent representations, i.e. $\vec{i_j} \in \mathbb{R}^{m_i}$, $\vec{u_j} \in \mathbb{R}^{m_u}$ and $\vec{d_j} \in \mathbb{R}^{m_d}$ respectively.

\minisec{Domain Contextualization} In order to transform item embeddings from multiple domains to a shared latent space, we apply a dense transformation layer to the concatenation of the item embedding and its associated domain embedding as follows:
\begin{equation}
    \vec{i_j}' = \mathbf{W_d} \cdot (\vec{i_j} \;\|\; \vec{d_j})
\end{equation}

Where $\|$ is the concatenation operator and $\mathbf{W_d}$ is a matrix of learnable parameters. Note that other aggregation operators could also be used, such as element-wise summation or Hadamard product.

\minisec{Recent Interest Extractor} This block receives the selected graph representation of the user behavior history $\mathcal{G}_u$, already contextualized by the previous module, and it is responsible for temporally aggregating implicit signals into a strong recent user interest representation. This process is performed in two steps:

\minipara{Domain-Aware Gated Graph Convolution} To learn inter-domain sequential patterns and intra-domain structural constraints, it is important to model domain commonalities by considering how much information a domain is able to pass onto another domain. Therefore, an edge weight is computed using the source and target domains as edge attributes:
    
\begin{equation}
    e_{z, j} = \langle \mathbf{W_{src}} \cdot \vec{d_z}, \mathbf{W_{trg}} \cdot \vec{d_j} \rangle
\end{equation}

Where $\mathbf{W_{src}}$ and $\mathbf{W_{trg}}$ are learnable parameters and $\langle \vec{x}, \vec{y} \rangle$ is the dot product operator. 
The next step is to learn smooth node representations with a proper sequence modeling technique applied to graphs. We opted for stacking $L$ Gated Graph Convolutional (GGCN) layers \cite{ggcn} with the previously computed cross-domain edge weights:

\begin{equation}
    \vec{h}^{(l+1)}_{j} = GGCN(e_{z, j}, \vec{h}^{(l)}_z)
\end{equation}

Where $\vec{h}^{(0)}_z = \vec{i_j}'$.
    
\minipara{Attention Graph Readout} A graph level embedding of the short-term user interests is obtained by a weighted aggregation of the node embeddings. To compute those weights, a local attention mechanism is applied to each node $j$ and the node corresponding to the last interaction of $\mathcal{B}_u$:
    
\begin{equation*}
    \alpha_j = softmax_j\left (\mathbf{W_{l_2}} \cdot \sigma(\mathbf{W_{l_1}} \cdot (\vec{h}^{(L)}_j\;\|\; \vec{h}^{(L)}_N))^T\right)
\end{equation*}

Where $\mathbf{W_{l_1}}$  and $\mathbf{W_{l_2}}$ are learnable parameters and $\sigma$ is a non-linear activation function. Finally, a weighted aggregation with the computed coefficients generates the graph embedding:

\begin{equation}
    \vec{r}_u = \sum_{j \in \mathcal{V}_u} \alpha_j \cdot \vec{h}^{(l+1)}_j
\end{equation}

\minisec{Global Interest Extractor} A graph built from  user history is useful for extracting strong recent interests but the history may also contain noisy interactions that confound long-term core user preferences. The global interest extractor branch is tasked with obtaining a global time independent user interest representation. It consists of the following modules:

\minipara{Graph Structure Learning} A new graph representation $\mathcal{G}_u' = \{\mathcal{V}_u, \mathcal{E}_u'\}$ whose adjacency matrix $\mathcal{A}'$ is created based on a relative ranking of multi-head kernel similarity scores between any two vertices in $\mathcal{V}_u$:
    
\begin{equation}
    \mathcal{A}^h_{j, z} = Kernel(\mathbf{W^h_{gsl}}\cdot \vec{i_j}', \mathbf{W^h_{gsl}}\cdot \vec{i_z}')
\end{equation}

\begin{equation}
    \mathcal{A}_{j, z} = \frac{1}{H} \sum^{H}_{h=1} \mathcal{A}^h_{j, z}
\end{equation}
    
Where $H$ denotes the number of kernel heads to extract specific node similarities and $\mathbf{W^h_{gsl}}$ are learnable parameters. We select the cosine similarity as the kernel function.
Following \cite{surge}, a relative ranking with a threshold $\gamma$ equal to 0.5 is applied for graph sparsification:

\begin{equation}
    \mathcal{A}_{j, z}' = \begin{cases}
    1, & \text{if $\mathcal{A}_{j, z} \geq \gamma$};\\
    0, & \text{otherwise};
    \end{cases}
\end{equation}
    
\minipara{Local Graph Pooling} MemPooling \cite{memgnn}, a hierarchical graph representation learning technique based on multiple memory layers for local clustering, is applied on top of a Graph Attention (GAT) query network to obtain a global user interest embedding of the last $N$ interactions:
    
\begin{equation}
    \vec{g}_u = MemPool\left(GAT({\vec{i_j}'|j \in \mathcal{G}_u'})\right)
\end{equation}
    
\minipara{User Contextualization} Given that a user can have more than $N$ past interactions, a ``residual" user contextualization is aimed at a more global interest representation:
    
\begin{equation}
    \vec{g_u}' = \mathbf{W_u} \cdot \left( \vec{g}_u \;\|\; \vec{u} \right)
\end{equation}

Where $\mathbf{W_u}$ are learnable parameters.

Note that learning a new graph structure based on item similarities increases the model cross-domain representation capabilities as it helps to pull similar items from different domains together in the shared space.

\minisec{Classification Network} Previous MDR work has focused their efforts on this block by extracting ideas from MTL strategies combining domain-shared and domain-specific parameters. However, given the aim of this paper on multi-domain data modeling, we decided to implement a single tower network with global batch normalization and FCN. This tower network receives the concatenation of the contextualized candidate item embedding and both, recent and global user interest representations $(\vec{i_k}, \vec{g_u}', \vec{r_u})$, and it outputs the probability of the candidate item being clicked next by the current user $(\hat{y}_k)$. It is important to note that we leave the combination of MDR-MTL strategies and our MDR graph-based representational model for future work.

%% file: discussion.tex
\section{Experiments}

\minisection{Dataset\footnote{Code and dataset partitions are available at \url{https://github.com/alarca94/magrec}.}} Following previous work \cite{conet,dasl,mamdr}, we use the Amazon 5-core review dataset \cite{amazon_dataset} by combining data corresponding to different domains (product categories) with varying degrees of users overlap, dataset size and click-trough-rates (CTR). In order to bridge CDR and MDR evaluation scenarios, we form eight sub-datasets: four 2-domain,
 one 3-domain, one 6-domain and one 13-domain partition. Similar to previous studies \cite{mamdr}, for each partition, we keep the existing user-item reviews as positive samples and perform negative sampling on the set of items that the user has not interacted with, using a randomly generated CTR per domain to simulate domain distinction. For sequential recommendation, we perform a sliding window approach on the user histories and set the minimum and maximum window lengths to 5 and 80 respectively. Finally, for each user, a temporal split of ratio 6:2:2 is used to create the train, validation and test sets.
Table \ref{tab:data_stats} summarizes the basic statistics for all datasets.

\begin{table}[htb!]
\footnotesize
    \centering
    \caption{Dataset statistics.}
    \resizebox{0.6\columnwidth}{!}{
    \begin{tabular}{|l|c|c|c|c|c|}
        \toprule
        Dataset & \# User & \# Item & \# Train & \# Val & \# Test \\
        \midrule
        Amazon-2a & 153,658 & 55,089 & 1,766,478 & 267,805 & 267,884 \\
        Amazon-2b & 290,631 & 105,609 & 3,593,621 & 610,585 & 610,537 \\
        Amazon-2c & 188,539 & 59,978 & 2,245,655 & 348,101 & 347,917 \\
        Amazon-2d & 254,736 & 95,394 & 3,009,960 & 555,399 & 555,701 \\
        Amazon-3 & 113,144 & 33,193 & 1,443,804 & 220,155 & 220,109 \\
        Amazon-6 & 444,737 & 170,815 & 7,580,825 & 1,397,290 & 1,395,688 \\
        Amazon-13 & 500,569 & 212,241 & 7,105,872 & 1,380,084 & 1,379,630 \\
        \bottomrule
    \end{tabular}
    }
    \label{tab:data_stats}
\end{table}

\minisection{Baselines} To demonstrate the sequential modeling effectiveness of MAGRec, we compare it against competitive single-domain recommenders, including non-sequential, sequential and graph-based alternatives. Furthermore, we consider several MTL and MDR strong baselines to determine how well our model performs in multi-domain settings:

\begin{itemize}
    \item WDL \cite{wdl}: non-sequential model based on the combination of wide linear and deep neural models with cross-product feature transformations.
    \item DIN \cite{din}: non-sequential model that aggregates the user behavior history using a softmax attention pooling based on the candidate item.
    \item DIEN \cite{dien}: sequential model with a two-layer GRU that implements an attentional update gate for a proper interest evolution modeling.
    \item FGNN \cite{fgnn}: competitive session-based recommender model adapted to predict CTR. It uses several weighted attention graph layers followed by a GRU set-to-set readout function.
    \item Shared-Bottom (SB) \cite{sharedbottom}: MTL model with shared parameters in the bottom layers and $|\mathcal{D}|$ domain-specific tower networks.
    \item MMoE \cite{mmoe}: MTL model that adopts a Mixture of Experts layout with $|\mathcal{D}|$ experts and a gating mechanism per domain connecting experts to the respective tower networks.
    \item STAR \cite{star}: MDR model with partitioned normalization and star topology network to leverage shared and specific domain knowledge. They also use an auxiliary domain-aware network.
    \item MAMDR \cite{mamdr}: MDR learning method that uses domain regularization and domain negotiation on top of a star topology network.
\end{itemize}

To give a fair comparison, all non-sequential models are fed with a mean pooled representation of the user behavior sequence. Some of the methods can be used as a complementary to ours.

\minisection{Evaluation Metrics} We use logloss and area under the ROC curve (AUC) as it is the most common metrics used to evaluate the performance of CTR prediction. We report overall metrics for all domains and, wherever possible, domain-specific results.

\minisection{Implementation Details} Both graph-based models (FGNN\footnote{\url{https://github.com/RuihongQiu/FGNN}} and MAGRec) are implemented using Pytorch Geometric. MAMDR and STAR implementations correspond to the ones provided by MAMDR authors\footnote{\url{https://github.com/RManLuo/MAMDR}}. For the other methods, they have been implemented by DeepCTR \cite{deepctr}. To make a fair comparison, all FCNs are set to [128, 64] parameters; the user and item embedding sizes are 64; the domain embedding size is 128; the initial learning rate is set to $1e^{-3}$ and the batch size is 512; the selected optimizer is Adam and Binary Cross Entropy is the loss function. For the model specific parameters, in the case of DeepCTR and graph-based models, a TPE hyperparameter search is performed on the Amazon 2d dataset and the best hyperparameters are used for the rest of the datasets. MAGRec best hyperparameters are 2 GGCN layers, 1 GAT layer with 2 attention heads and 3 MemPooling layers with [32, 10, 1] centroids respectively.
Regarding MTL and MDR model specific parameters, we keep the ones selected in \cite{mamdr}. Experiments were conducted on a Tesla P100-PCIE-16GB GPU and 500GB RAM.

\minisection{Two Domains Results} The results for two domain datasets are presented in Table~\ref{tab:amazon2}. Overall and per-domain AUC and logloss values are presented. MAGRec outperforms all other methods in a meaning of single and combined domain performance for all scenarios.
Amazon-2c is the scenario with the biggest logloss difference between two domains. The second-best method in this setting is an MTL approach. MMoE is consistently better than others (on the second place after MAGRec). Single-domain recommenders in this scenario can still be strong baselines, i.e. DIN and DIEN outperform STAR in all datasets and MAMDR for Amazon-2{b,d}. Lastly, session-based FGNN performs the worst on all partitions.

\begin{table}[htb!]
\centering
\caption{Results on 2-domain datasets and different methods. Bold numbers indicate the leading results, while underlined numbers represent the second-best scores. Mean and std. from seven runs are presented.}
\resizebox{1\columnwidth}{!}{
\begin{tabular}{|cl|cc|cc|cc|}
\toprule
        Dataset  &    Model    & \multicolumn{2}{c|}{Domain 1} & \multicolumn{2}{c|}{Domain 2} & \multicolumn{2}{c|}{Both} \\
          &        &                                     logloss &                                         AUC &                                     logloss &                                         AUC &                                     logloss &                                         AUC \\
\midrule
    \parbox[t]{2mm}{\multirow{9}{*}{\rotatebox[origin=c]{90}{Amazon-2a}}} & DIEN &  0.510 \scriptsize{$\pm$ 0.007} &  0.786 \scriptsize{$\pm$ 0.003} &  0.393 \scriptsize{$\pm$ 0.003} &  0.853 \scriptsize{$\pm$ 0.003} &  0.452 \scriptsize{$\pm$ 0.004} &  0.819 \scriptsize{$\pm$ 0.002} \\
          & DIN &  0.496 \scriptsize{$\pm$ 0.003} &  0.806 \scriptsize{$\pm$ 0.002} &  0.398 \scriptsize{$\pm$ 0.004} &  0.842 \scriptsize{$\pm$ 0.002} &  0.447 \scriptsize{$\pm$ 0.004} &  0.824 \scriptsize{$\pm$ 0.001} \\
          & FGNN &  0.530 \scriptsize{$\pm$ 0.003} &  0.766 \scriptsize{$\pm$ 0.002} &  0.451 \scriptsize{$\pm$ 0.003} &  0.792 \scriptsize{$\pm$ 0.004} &  0.491 \scriptsize{$\pm$ 0.003} &  0.779 \scriptsize{$\pm$ 0.003} \\
          & MAMDR &  0.481 \scriptsize{$\pm$ 0.007} &  0.819 \scriptsize{$\pm$ 0.005} &  0.377 \scriptsize{$\pm$ 0.003} &  0.865 \scriptsize{$\pm$ 0.002} &  0.429 \scriptsize{$\pm$ 0.004} &  0.842 \scriptsize{$\pm$ 0.002} \\
          & MMoE &  0.474 \scriptsize{$\pm$ 0.013} &  0.823 \scriptsize{$\pm$ 0.007} &  \underline{0.372 \scriptsize{$\pm$ 0.009}} &  \underline{0.869 \scriptsize{$\pm$ 0.005}} &  \underline{0.423 \scriptsize{$\pm$ 0.003}} &  \underline{0.846 \scriptsize{$\pm$ 0.001}} \\
          & SB &  \underline{0.470 \scriptsize{$\pm$ 0.013}} &  \underline{0.823 \scriptsize{$\pm$ 0.010}} &  0.375 \scriptsize{$\pm$ 0.006} &  0.865 \scriptsize{$\pm$ 0.004} &  0.423 \scriptsize{$\pm$ 0.005} &  0.844 \scriptsize{$\pm$ 0.004} \\
          & STAR &  0.543 \scriptsize{$\pm$ 0.032} &  0.777 \scriptsize{$\pm$ 0.005} &  0.419 \scriptsize{$\pm$ 0.021} &  0.830 \scriptsize{$\pm$ 0.008} &  0.481 \scriptsize{$\pm$ 0.022} &  0.804 \scriptsize{$\pm$ 0.003} \\
          & WDL &  0.481 \scriptsize{$\pm$ 0.013} &  0.816 \scriptsize{$\pm$ 0.008} &  0.378 \scriptsize{$\pm$ 0.005} &  0.863 \scriptsize{$\pm$ 0.002} &  0.430 \scriptsize{$\pm$ 0.004} &  0.840 \scriptsize{$\pm$ 0.003} \\
          & MAGRec &  \textbf{0.466 \scriptsize{$\pm$ 0.004}} &  \textbf{0.831 \scriptsize{$\pm$ 0.001}} &  \textbf{0.363 \scriptsize{$\pm$ 0.003}} &  \textbf{0.875 \scriptsize{$\pm$ 0.001}} &  \textbf{0.415 \scriptsize{$\pm$ 0.003}} &  \textbf{0.853 \scriptsize{$\pm$ 0.001}} \\
\midrule
    \parbox[t]{2mm}{\multirow{9}{*}{\rotatebox[origin=c]{90}{Amazon-2b}}} & DIEN &  \textbf{0.407 \scriptsize{$\pm$ 0.055}} &  0.849 \scriptsize{$\pm$ 0.004} &  0.405 \scriptsize{$\pm$ 0.022} &  0.841 \scriptsize{$\pm$ 0.001} &  \underline{0.406 \scriptsize{$\pm$ 0.033}} &  0.845 \scriptsize{$\pm$ 0.003} \\
          & DIN &  0.451 \scriptsize{$\pm$ 0.000} &  0.845 \scriptsize{$\pm$ 0.000} &  0.400 \scriptsize{$\pm$ 0.000} &  0.838 \scriptsize{$\pm$ 0.000} &  0.425 \scriptsize{$\pm$ 0.000} &  0.841 \scriptsize{$\pm$ 0.000} \\
          & FGNN &  0.485 \scriptsize{$\pm$ 0.006} &  0.820 \scriptsize{$\pm$ 0.005} &  0.421 \scriptsize{$\pm$ 0.003} &  0.822 \scriptsize{$\pm$ 0.003} &  0.453 \scriptsize{$\pm$ 0.005} &  0.821 \scriptsize{$\pm$ 0.004} \\
          & MAMDR &  0.467 \scriptsize{$\pm$ 0.007} &  0.834 \scriptsize{$\pm$ 0.004} &  0.408 \scriptsize{$\pm$ 0.004} &  0.836 \scriptsize{$\pm$ 0.002} &  0.437 \scriptsize{$\pm$ 0.005} &  0.835 \scriptsize{$\pm$ 0.003} \\
          & MMoE &  0.443 \scriptsize{$\pm$ 0.027} &  \underline{0.851 \scriptsize{$\pm$ 0.016}} &  \underline{0.386 \scriptsize{$\pm$ 0.017}} &  \underline{0.849 \scriptsize{$\pm$ 0.011}} &  0.414 \scriptsize{$\pm$ 0.015} &  \underline{0.850 \scriptsize{$\pm$ 0.010}} \\
          & SB &  0.472 \scriptsize{$\pm$ 0.031} &  0.830 \scriptsize{$\pm$ 0.018} &  0.393 \scriptsize{$\pm$ 0.010} &  0.841 \scriptsize{$\pm$ 0.010} &  0.433 \scriptsize{$\pm$ 0.020} &  0.836 \scriptsize{$\pm$ 0.013} \\
          & STAR &  0.467 \scriptsize{$\pm$ 0.010} &  0.832 \scriptsize{$\pm$ 0.003} &  0.417 \scriptsize{$\pm$ 0.019} &  0.828 \scriptsize{$\pm$ 0.004} &  0.442 \scriptsize{$\pm$ 0.007} &  0.830 \scriptsize{$\pm$ 0.001} \\
          & WDL &  0.461 \scriptsize{$\pm$ 0.019} &  0.838 \scriptsize{$\pm$ 0.014} &  0.396 \scriptsize{$\pm$ 0.013} &  0.842 \scriptsize{$\pm$ 0.008} &  0.428 \scriptsize{$\pm$ 0.009} &  0.840 \scriptsize{$\pm$ 0.008} \\
          & MAGRec &  \underline{0.416 \scriptsize{$\pm$ 0.002}} &  \textbf{0.869 \scriptsize{$\pm$ 0.002}} &  \textbf{0.378 \scriptsize{$\pm$ 0.003}} &  \textbf{0.856 \scriptsize{$\pm$ 0.002}} &  \textbf{0.397 \scriptsize{$\pm$ 0.002}} &  \textbf{0.862 \scriptsize{$\pm$ 0.002}} \\
\midrule
    \parbox[t]{2mm}{\multirow{9}{*}{\rotatebox[origin=c]{90}{Amazon-2c}}} & DIEN &  \textbf{0.446 \scriptsize{$\pm$ 0.060}} &  0.805 \scriptsize{$\pm$ 0.006} &  0.411 \scriptsize{$\pm$ 0.023} &  0.835 \scriptsize{$\pm$ 0.002} &  0.429 \scriptsize{$\pm$ 0.036} &  0.820 \scriptsize{$\pm$ 0.004} \\
          & DIN &  0.496 \scriptsize{$\pm$ 0.001} &  0.805 \scriptsize{$\pm$ 0.001} &  0.399 \scriptsize{$\pm$ 0.001} &  0.834 \scriptsize{$\pm$ 0.001} &  0.448 \scriptsize{$\pm$ 0.001} &  0.819 \scriptsize{$\pm$ 0.001} \\
          & FGNN &  0.525 \scriptsize{$\pm$ 0.002} &  0.773 \scriptsize{$\pm$ 0.000} &  0.432 \scriptsize{$\pm$ 0.001} &  0.808 \scriptsize{$\pm$ 0.001} &  0.478 \scriptsize{$\pm$ 0.001} &  0.791 \scriptsize{$\pm$ 0.001} \\
          & MAMDR &  0.484 \scriptsize{$\pm$ 0.005} &  0.816 \scriptsize{$\pm$ 0.004} &  0.396 \scriptsize{$\pm$ 0.002} &  0.844 \scriptsize{$\pm$ 0.001} &  0.440 \scriptsize{$\pm$ 0.002} &  0.830 \scriptsize{$\pm$ 0.002} \\
          & MMoE &  0.462 \scriptsize{$\pm$ 0.009} &  \underline{0.829 \scriptsize{$\pm$ 0.006}} &  \underline{0.393 \scriptsize{$\pm$ 0.007}} &  \underline{0.845 \scriptsize{$\pm$ 0.005}} &  \underline{0.428 \scriptsize{$\pm$ 0.005}} &  \underline{0.837 \scriptsize{$\pm$ 0.003}} \\
          & SB &  0.468 \scriptsize{$\pm$ 0.011} &  0.825 \scriptsize{$\pm$ 0.007} &  0.394 \scriptsize{$\pm$ 0.007} &  0.845 \scriptsize{$\pm$ 0.006} &  0.431 \scriptsize{$\pm$ 0.007} &  0.835 \scriptsize{$\pm$ 0.005} \\
          & STAR &  0.528 \scriptsize{$\pm$ 0.017} &  0.782 \scriptsize{$\pm$ 0.004} &  0.411 \scriptsize{$\pm$ 0.008} &  0.827 \scriptsize{$\pm$ 0.003} &  0.469 \scriptsize{$\pm$ 0.006} &  0.805 \scriptsize{$\pm$ 0.001} \\
          & WDL &  0.469 \scriptsize{$\pm$ 0.004} &  0.822 \scriptsize{$\pm$ 0.004} &  0.401 \scriptsize{$\pm$ 0.007} &  0.841 \scriptsize{$\pm$ 0.006} &  0.435 \scriptsize{$\pm$ 0.005} &  0.831 \scriptsize{$\pm$ 0.004} \\
          & MAGRec &  \underline{0.462 \scriptsize{$\pm$ 0.007}} &  \textbf{0.830 \scriptsize{$\pm$ 0.006}} &  \textbf{0.383 \scriptsize{$\pm$ 0.004}} &  \textbf{0.852 \scriptsize{$\pm$ 0.004}} &  \textbf{0.423 \scriptsize{$\pm$ 0.006}} &  \textbf{0.841 \scriptsize{$\pm$ 0.005}} \\
\midrule
    \parbox[t]{2mm}{\multirow{9}{*}{\rotatebox[origin=c]{90}{Amazon-2d}}} & DIEN &  0.427 \scriptsize{$\pm$ 0.026} &  0.857 \scriptsize{$\pm$ 0.005} &  0.413 \scriptsize{$\pm$ 0.029} &  0.844 \scriptsize{$\pm$ 0.005} &  0.420 \scriptsize{$\pm$ 0.023} &  0.851 \scriptsize{$\pm$ 0.004} \\
          & DIN &  0.413 \scriptsize{$\pm$ 0.008} &  0.872 \scriptsize{$\pm$ 0.005} &  0.398 \scriptsize{$\pm$ 0.006} &  0.842 \scriptsize{$\pm$ 0.002} &  0.405 \scriptsize{$\pm$ 0.007} &  0.857 \scriptsize{$\pm$ 0.003} \\
          & FGNN &  0.468 \scriptsize{$\pm$ 0.004} &  0.837 \scriptsize{$\pm$ 0.002} &  0.427 \scriptsize{$\pm$ 0.003} &  0.825 \scriptsize{$\pm$ 0.002} &  0.448 \scriptsize{$\pm$ 0.003} &  0.831 \scriptsize{$\pm$ 0.002} \\
          & MAMDR &  0.473 \scriptsize{$\pm$ 0.017} &  0.857 \scriptsize{$\pm$ 0.005} &  \underline{0.391 \scriptsize{$\pm$ 0.001}} &  \underline{0.852 \scriptsize{$\pm$ 0.001}} &  0.432 \scriptsize{$\pm$ 0.008} &  0.855 \scriptsize{$\pm$ 0.002} \\
          & MMoE &  \underline{0.393 \scriptsize{$\pm$ 0.014}} &  \underline{0.885 \scriptsize{$\pm$ 0.009}} &  0.394 \scriptsize{$\pm$ 0.016} &  0.849 \scriptsize{$\pm$ 0.012} &  \underline{0.393 \scriptsize{$\pm$ 0.014}} &  \underline{0.867 \scriptsize{$\pm$ 0.010}} \\
          & SB &  0.416 \scriptsize{$\pm$ 0.021} &  0.870 \scriptsize{$\pm$ 0.015} &  0.408 \scriptsize{$\pm$ 0.015} &  0.837 \scriptsize{$\pm$ 0.013} &  0.412 \scriptsize{$\pm$ 0.017} &  0.853 \scriptsize{$\pm$ 0.013} \\
          & STAR &  0.455 \scriptsize{$\pm$ 0.006} &  0.842 \scriptsize{$\pm$ 0.006} &  0.431 \scriptsize{$\pm$ 0.014} &  0.824 \scriptsize{$\pm$ 0.005} &  0.443 \scriptsize{$\pm$ 0.005} &  0.833 \scriptsize{$\pm$ 0.002} \\
          & WDL &  0.414 \scriptsize{$\pm$ 0.011} &  0.871 \scriptsize{$\pm$ 0.008} &  0.418 \scriptsize{$\pm$ 0.019} &  0.830 \scriptsize{$\pm$ 0.014} &  0.416 \scriptsize{$\pm$ 0.015} &  0.850 \scriptsize{$\pm$ 0.011} \\
          & MAGRec &  \textbf{0.375 \scriptsize{$\pm$ 0.002}} &  \textbf{0.897 \scriptsize{$\pm$ 0.001}} &  \textbf{0.363 \scriptsize{$\pm$ 0.001}} &  \textbf{0.873 \scriptsize{$\pm$ 0.001}} &  \textbf{0.369 \scriptsize{$\pm$ 0.001}} &  \textbf{0.885 \scriptsize{$\pm$ 0.001}} \\
\bottomrule
\end{tabular}
}
\label{tab:amazon2}
\end{table}

\minisection{Input representation analysis} As described in the \emph{Methodology} section,
an input graph for MAGRec can be prepared in a few different ways. Consequently, we have experimented with three different graph representations i.e. \textit{Disjoint}, \textit{Flattened} and \textit{Interacting History}. In \textit{Disjoint History}, domains are considered as independent data sources, meaning that for a candidate item $i_k$ from domain $d_k$, the graph is constructed from the past user history in that particular domain. This representation is inline with the way MDR models \cite{mamdr,star} handle alternate domain training and, therefore, it is not able to exploit the existing cross-domain behaviors in the overall user history. A complete but naive cross-domain representation is the sequence of user-item interactions in all domains as a single timeline, thus, filling the gap between specific domain user sessions with the user sessions occurring in other domains. This representation is what we refer to as \textit{Flattened History}. The downside of it is that user interests are assumed to evolve smoothly across all domains and unrelated domains could potentially create an interest bottleneck in a k-hop message passing for user recent interest modelling. To overcome this limitation, we propose an \textit{Interacting History} representation in which the \textit{Flattened History} is enriched with domain skip connections to enable uninterrupted intra-domain paths and proper signal propagation in cases where two domains share little to no commonalities. Table~\ref{tab:graph} compares  all three representations for two domain setting on Amazon-2a dataset and different methods. It is clear that \emph{Interacting History} gives the best results with $26.5\%$ and $0.5\%$ improvement over \emph{Disjoint} and \emph{Flattened History} representations respectively for MAGRec. Results are consistent for sequential (DIN, DIEN) and graph-based models (FGNN, MAGRec) proving the importance of multi-domain context as well as intra- and inter-domain paths. FGNN cannot benefit from the \emph{Interacting History}, as it lacks a dedicated mechanism for Domain Contextualization during message propagation.

\begin{table}[tb]
\centering
\caption{Comparison of different methods for graph $\mathcal{G}_u$ preparation on Amazon-2a dataset for selected methods. DIN and DIEN cannot be combined with \emph{Interacting}.}
\resizebox{1.0\columnwidth}{!}{

\begin{tabular}{|l|cc|cc|cc|cc|}
\toprule
      Input & 
      \multicolumn{2}{c|}{DIN} & \multicolumn{2}{c|}{DIEN} &
      \multicolumn{2}{c|}{FGNN} & \multicolumn{2}{c|}{MAGRec} \\
            &  
            logloss &      AUC &
            logloss &      AUC &
            logloss &      AUC &  
            logloss &      AUC \\
\midrule
Disjoint & 0.486 \scriptsize{$\pm$ 0.012} & 0.778 \scriptsize{$\pm$ 0.017} & 0.493 \scriptsize{$\pm$ 0.013} & 0.790 \scriptsize{$\pm$ 0.003} & 0.496 \scriptsize{$\pm$ 0.001} & 0.774 \scriptsize{$\pm$ 0.004} & 0.565 \scriptsize{$\pm$ 0.209} & 0.827 \scriptsize{$\pm$ 0.021} \\
 Flattened & \textbf{0.447 \scriptsize{$\pm$ 0.004}} & \textbf{0.824 \scriptsize{$\pm$ 0.001}} & \textbf{0.453 \scriptsize{$\pm$ 0.004}} & \textbf{0.816 \scriptsize{$\pm$ 0.002}} & \textbf{0.495 \scriptsize{$\pm$ 0.003}} & \textbf{0.779 \scriptsize{$\pm$ 0.003}} & 0.417 \scriptsize{$\pm$ 0.003} & 0.849 \scriptsize{$\pm$ 0.002} \\
 Interacting &    \multicolumn{1}{c}{-}    &    -    &    -    &    -    & 0.497 \scriptsize{$\pm$ 0.002} & 0.778 \scriptsize{$\pm$ 0.002} & \textbf{0.415 \scriptsize{$\pm$ 0.003}} & \textbf{0.853 \scriptsize{$\pm$ 0.001}} \\
\bottomrule
\end{tabular}
}

\label{tab:graph}
\end{table}

\begin{figure}[tb]
\centering
\includegraphics[width=.8\columnwidth,keepaspectratio]{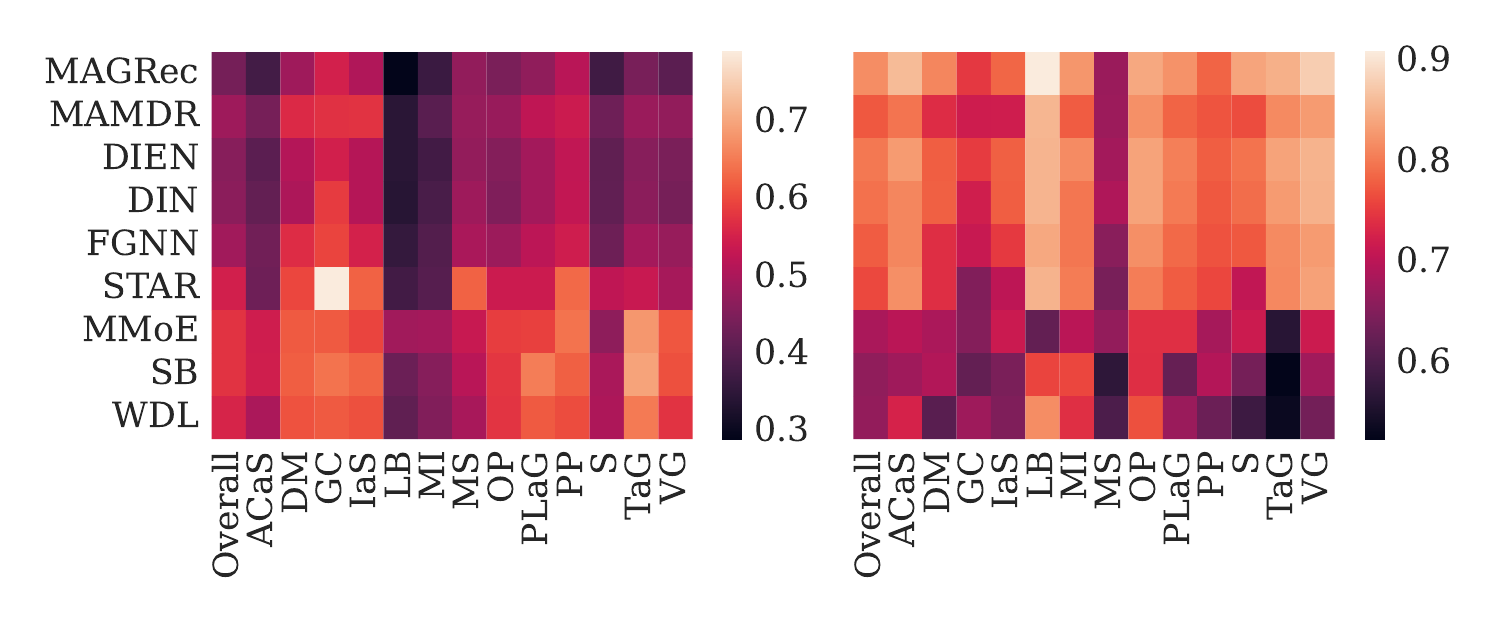}
\caption{Results for all 13 domains and methods presented as a heatmap. Logloss (left) and AUC (right).}
\label{fig:hm_13}
\end{figure}

\minisection{More domains scenarios} In order to evaluate our method performance beyond two domain scenarios, we use three, six and 13 domains. The results are in Table~\ref{tab:more}. MAGRec receives the best logloss and AUC in all scenarios. Interestingly, SDR models (i.e., DIN, DIEN) achieve second-best performance on 13-domains where both, MAMDR and STAR, sttrugle to model all domains well at the same time. As stated in \cite{mamdr}, the incremental drop in performance is even more noticeable for STAR. MTL-based models such as MMoE also cope to handle larger number of domains. Figure~\ref{fig:hm_13} presents variability across all domains and used methods. MAGRec and DIEN are clear winners here. While MMoE, SB, WDL are worse, with a high variability between domains.

\begin{table}[tb]
\centering
\caption{Average logloss and AUC values for different methods and many domains scenarios}
\resizebox{0.95\columnwidth}{!}{

\begin{tabular}{|l|cc|cc|cc|}
\toprule
{Model} & \multicolumn{2}{c|}{3 domains} & \multicolumn{2}{c|}{6 domains} & \multicolumn{2}{c|}{13 domains} \\
{} &                                     logloss &                                         AUC &                                     logloss &                                         AUC &                                     logloss &                                         AUC \\
\midrule
SB         &  0.482 \scriptsize{$\pm$ 0.007} &  0.802 \scriptsize{$\pm$ 0.007} &  0.516 \scriptsize{$\pm$ 0.017} &  0.736 \scriptsize{$\pm$ 0.023} &  0.574 \scriptsize{$\pm$ 0.022} &  0.661 \scriptsize{$\pm$ 0.016} \\
MMoE       &  0.489 \scriptsize{$\pm$ 0.037} &  0.809 \scriptsize{$\pm$ 0.003} &  0.514 \scriptsize{$\pm$ 0.029} &  0.754 \scriptsize{$\pm$ 0.022} &  0.572 \scriptsize{$\pm$ 0.026} &  0.683 \scriptsize{$\pm$ 0.009} \\
WDL        &  0.478 \scriptsize{$\pm$ 0.008} &  0.804 \scriptsize{$\pm$ 0.006} &  0.505 \scriptsize{$\pm$ 0.007} &  0.748 \scriptsize{$\pm$ 0.013} &  0.553 \scriptsize{$\pm$ 0.011} &  0.663 \scriptsize{$\pm$ 0.014} \\
STAR       &  0.528 \scriptsize{$\pm$ 0.020} &  0.766 \scriptsize{$\pm$ 0.003} &  0.488 \scriptsize{$\pm$ 0.014} &  0.810 \scriptsize{$\pm$ 0.002} &  0.545 \scriptsize{$\pm$ 0.024} &  0.760 \scriptsize{$\pm$ 0.003} \\
FGNN       &  0.512 \scriptsize{$\pm$ 0.001} &  0.766 \scriptsize{$\pm$ 0.001} &  0.453 \scriptsize{$\pm$ 0.001} &  0.808 \scriptsize{$\pm$ 0.002} &  0.486 \scriptsize{$\pm$ 0.002} &  0.775 \scriptsize{$\pm$ 0.000} \\
MAMDR      &  \underline{0.472 \scriptsize{$\pm$ 0.002}} &  \underline{0.812 \scriptsize{$\pm$ 0.002}} &  0.430 \scriptsize{$\pm$ 0.004} &  \underline{0.835 \scriptsize{$\pm$ 0.001}} &  0.483 \scriptsize{$\pm$ 0.036} &  0.772 \scriptsize{$\pm$ 0.042} \\
DIN        &  0.473 \scriptsize{$\pm$ 0.000} &  0.804 \scriptsize{$\pm$ 0.000} &  \underline{0.427 \scriptsize{$\pm$ 0.000}} &  0.818 \scriptsize{$\pm$ 0.000} &  0.461 \scriptsize{$\pm$ 0.000} &  0.791 \scriptsize{$\pm$ 0.000} \\
DIEN       &  0.475 \scriptsize{$\pm$ 0.004} &  0.807 \scriptsize{$\pm$ 0.002} &  0.427 \scriptsize{$\pm$ 0.001} &  0.825 \scriptsize{$\pm$ 0.001} &  \underline{0.457 \scriptsize{$\pm$ 0.005}} &  \underline{0.797 \scriptsize{$\pm$ 0.001}} \\
MAGRec     &  \textbf{0.461 \scriptsize{$\pm$ 0.003}} &  \textbf{0.819 \scriptsize{$\pm$ 0.003}} &  \textbf{0.416 \scriptsize{$\pm$ 0.001}} &  \textbf{0.836 \scriptsize{$\pm$ 0.001}} &  \textbf{0.436 \scriptsize{$\pm$ 0.001}} &  \textbf{0.815 \scriptsize{$\pm$ 0.002}} \\
\bottomrule
\end{tabular}
}
\label{tab:more}
\end{table}

\minisection{Ablation Study} In Table~\ref{tab:ablation_2a}, we present the ablation study results for MAGRec network, where particular modules of the network are disabled. 
The biggest improvement in logloss $0.08$ is when Recent Interest Extractor (RIE) is used. Then, Global Interest Extractor (GIE) gives a smaller boost of $0.002$. When combining all, with Domain Contextualization (DC), we observe the best outcome of $0.369$ logloss value. Combining DC with GIE alone does not help to get better overall performance.

\begin{table}[htb!]
\centering
\caption{Results for MAGRec and Amazon-2d dataset where Recent Interest Extractor (RIE), Global Interest Extraction (GIE) and Domain Contextualization (DC) are ablated.
}
\resizebox{0.98\columnwidth}{!}{
\begin{tabular}{|ccc|cc|cc|cc|}
\toprule
  RIE &   GIE &    DC & \multicolumn{2}{c|}{Domain 1} & \multicolumn{2}{c|}{Domain 2} & \multicolumn{2}{c|}{Both} \\
   &    &    &                        logloss &                            AUC &                        logloss &                            AUC &                        logloss$\downarrow$ &                            AUC \\
\midrule
  & \checkmark & \checkmark & 0.434 \scriptsize{$\pm$ 0.001} & 0.805 \scriptsize{$\pm$ 0.002} & 0.487 \scriptsize{$\pm$ 0.001} & 0.812 \scriptsize{$\pm$ 0.003} & 0.460 \scriptsize{$\pm$ 0.001} & 0.808 \scriptsize{$\pm$ 0.002} \\
  & \checkmark &   & 0.429 \scriptsize{$\pm$ 0.001} & 0.810 \scriptsize{$\pm$ 0.001} & 0.477 \scriptsize{$\pm$ 0.003} & 0.820 \scriptsize{$\pm$ 0.002} & 0.453 \scriptsize{$\pm$ 0.002} & 0.815 \scriptsize{$\pm$ 0.001} \\
\checkmark &   &   & 0.365 \scriptsize{$\pm$ 0.001} & 0.872 \scriptsize{$\pm$ 0.001} & 0.381 \scriptsize{$\pm$ 0.002} & 0.893 \scriptsize{$\pm$ 0.001} & 0.373 \scriptsize{$\pm$ 0.001} & 0.882 \scriptsize{$\pm$ 0.000} \\
\checkmark & \checkmark &   & 0.364 \scriptsize{$\pm$ 0.000} & 0.873 \scriptsize{$\pm$ 0.001} & 0.378 \scriptsize{$\pm$ 0.000} & 0.895 \scriptsize{$\pm$ 0.000} & 0.371 \scriptsize{$\pm$ 0.000} & 0.884 \scriptsize{$\pm$ 0.001} \\
\checkmark &   & \checkmark & 0.364 \scriptsize{$\pm$ 0.002} & 0.874 \scriptsize{$\pm$ 0.002} & 0.376 \scriptsize{$\pm$ 0.003} & 0.897 \scriptsize{$\pm$ 0.001} & 0.370 \scriptsize{$\pm$ 0.002} & 0.885 \scriptsize{$\pm$ 0.001} \\
\checkmark & \checkmark & \checkmark & 0.364 \scriptsize{$\pm$ 0.001} & 0.873 \scriptsize{$\pm$ 0.001} & 0.374 \scriptsize{$\pm$ 0.002} & 0.897 \scriptsize{$\pm$ 0.001} & \textbf{0.369 \scriptsize{$\pm$ 0.001}} & \textbf{0.885 \scriptsize{$\pm$ 0.001}} \\
\bottomrule
\end{tabular}
}
\label{tab:ablation_2a}
\end{table}

%% file: conclusions.tex
\section{Conclusions and Future Work}

This paper presents a new method, MAGRec, for multi-domain recommendation that uses GNNs to model intra- and inter-domain sequential relations.
In our experiments on publicly available datasets, our method shows better performance compared to state-of-the-art methods in multiple different settings: from two up to 13 different domain combinations. Additionally, we performed a series of experiments that proved the usefulness of \emph{Interacting History} representation as well as \emph{Recent} and \emph{Global Interest Extractors}.
As future work, higher-order graph representations could be explored as input for sparse multi-domain representation. Additionally, aforementioned integration with other MDR training strategies can further improve the results.